\documentclass[12pt,a4paper]{article}
\usepackage{amssymb}
\usepackage{graphicx}
\usepackage{amsmath}

\input{tcilatex}

\begin{document}

\title{Growth-optimal investments and numeraire portfolios under transaction
costs:\\
An analysis based on the von Neumann-Gale model$^{\ast }$}
\author{Wael Bahsoun$^{a}$, Igor V. Evstigneev$^{b}$ and Michael I. Taksar$%
^{c}$}
\date{\ \ \ }
\maketitle

\begin{abstract}
The aim of this work is to extend the capital growth theory developed by
Kelly, Breiman, Cover and others to asset market models with transaction
costs. We define a natural generalization of the notion of a numeraire
portfolio proposed by Long and show how such portfolios can be used for
constructing growth-optimal investment strategies. The analysis is based on
the classical von Neumann-Gale model of economic dynamics, a stochastic
version of which we use as a framework for the modelling of financial
markets with frictions.
\end{abstract}



{\small \noindent\textit{JEL-Classification:} C61, C62, O41, G10.}

{\small \noindent}\textbf{\rule{4.4in}{0.01in}}

{\small \noindent }$^{a}\,${\small School of Mathematics,} {\small %
Loughborough University, Leicestershire, LE11 3TU, UK.}

{\small \noindent }$^{b}\,${\small Economics Department, University of
Manchester, Oxford Road, Manchester M13 9PL, UK. E-mail:
igor.evstigneev@manchester.ac.uk. Phone: 0161-2754275. Fax: 0161-2754812.
(Corresponding author.)}

{\small \noindent }$^{c}\,${\small Mathematics Department, University of
Missouri, Columbia, MO 65211, USA.}

\pagebreak \textbf{\bigskip}

\section{Introduction}

How to invest in order to achieve the maximum growth rate of wealth in the
long run? This question has been in the focus of studies by Kelly \cite%
{Kelly1956}, Breiman \cite{Breiman1961}, Thorp \cite{Thorp1971}, Algoet and
Cover \cite{AlgoetCover1988}, Hakansson and Ziemba \cite{HakanssonZiemba1995}%
, Platen and Heath \cite{PlatenHeath2006}, and many others\footnote{%
To the list of those who contributed to this line of research, one has to
add the name of Claude Shannon---the famous founder of the mathematical
theory of information. Although he did not publish on investment-related
issues, his ideas expressed in his lectures on investment problems in the
1950s and 60s should be regarded as one of the main sources for that strand
of literature which we cite here. For the history of these ideas and the
related discussion see Cover \cite{Cover1998}.}. For the most part, results
available in the literature on capital growth pertain to markets without
transaction costs. Up to now, only some specialized models of markets with
friction have been analyzed in this field; see e.g. Taksar, Klass and Assaf 
\cite{TaksarKlassAssaf1988}, Iyengar and Cover \cite{IyengarCover2000},
Akian, Sulem and Taksar \cite{AkianSulemTaksar2001}, and Iyengar \cite%
{Iyengar2005}. The goal of the present work is to develop a capital growth
theory within a general discrete-time framework taking into account
proportional transaction costs. Our main tool in this study is one of the
fundamental models in mathematical economics---the von Neumann-Gale model of
economic growth.

The mathematical framework of the von Neumann-Gale model is a special class
of multivalued dynamical systems possessing certain properties of convexity
and homogeneity. The original theory of such systems (von Neumann \cite%
{vonNeumann1937}, Gale \cite{Gale1956}, Rockafellar \cite{Rockafellar1967},
Makarov and Rubinov \cite{MakarovRubinov1977}) aimed basically at the
modeling of economic dynamics. Initially, this theory was purely
deterministic; it did not reflect the influence of random factors on
economic growth. The importance of taking these factors into account was
realized early on. First attempts of constructing stochastic analogues of
the von Neumann--Gale model were undertaken in the 1970s by Dynkin \cite%
{Dynkin1971,Dynkin1972,DynkinYushkevich1979}, Radner \cite%
{Radner1971,Radner1972} and their research groups. However, the first attack
on the problem left many questions unanswered. Studies in this direction
faced serious mathematical difficulties. To overcome these difficulties, new
mathematical techniques were required, that were developed only during the
last decade---see \cite{EvstigneevSchenkHoppe2006,EvstigneevSchenkHoppe2008}
and \cite{BahsounEvstigneevTaksar2008}.

In a recent work of Dempster, Evstigneev and Taksar \cite%
{DempsterEvstigneevTaksar2006}, it has been observed that stochastic
analogues of von Neumann-Gale dynamical systems provide a natural and
convenient framework for the analysis of some fundamental problems in
finance (asset pricing and hedging under transaction costs). This paper
focuses on a different area of applications of such systems in finance. It
demonstrates how methods and concepts developed in the context of von
Neumann-Gale dynamics can be applied to the analysis of growth optimal
investments under transaction costs. A central notion related to von
Neumann-Gale dynamical systems, that of a \textit{rapid path}, plays a
crucial role in this work. We show that it yields a generalization of the
concept of a \textit{numeraire portfolio} (Long \cite{Long1990}) suitable
for the analysis of markets with transaction costs and trading constraints.
We obtain results on the existence of asymptotically optimal trading
strategies in markets with transaction costs by using results \cite%
{BahsounEvstigneevTaksar2008,EvstigneevSchenkHoppe2008} on the existence of
rapid paths in von Neumann-Gale systems.

The theory of von Neumann-Gale dynamical systems is one of the highlights of
mathematical economics. The results we refer to combine advanced methods of
ergodic theory, stochastic processes and functional analysis. In this paper,
we concentrate only on the modelling issues and the applications in finance.
The reader is referred to the literature cited for the proofs of the
mathematical results employed in this work. The main goal of this article is
to attract attention of theorists and practitioners working in quantitative
finance to new powerful methods developed in the field.

The paper is organized as follows. In Section 2 we describe the dynamic
securities market model we deal with. Section 3 introduces the basic
concepts and results related to the von Neumann-Gale dynamical systems. In
Section 4 we apply these results to the analysis of capital growth under
transaction costs. Section 5 concludes the paper.

\section{\textbf{Dynamic securities market model.}}

Let $s_{0},s_{1},...$ be a stochastic process with values in a measurable
space $S$. The process $(s_{t})_{t=0}^{+\infty}$ models random factors
influencing the market: the random element $s_{t}$ represents the ``state of
the world'' at date $t=0,1,...$. We denote by $%
s^{t}:=(s_{0},s_{1},...,s_{t}) $ the history of the process $(s_{t})$ up to
date $t$.

There are $n$ \textit{assets} traded in the market. A (\textit{contingent}) 
\textit{portfolio }of assets held by an investor at date $t$ is represented
by a vector 
\begin{equation*}
x_{t}(s^{t})=(x_{t}^{1}(s^{t}),...,x_{t}^{n}(s^{t}))
\end{equation*}
whose coordinates (portfolio positions) describe the holdings of assets $%
i=1,2,...,n$. The positions can be described either in terms of ``physical
units'' of assets or in terms of their market values. A contingent portfolio 
$x_{t}(s^{t})$ depends generally on the whole history $s^{t}$ of the process 
$(s_{t})$, which means that the investor can select his/her portfolio at
date $t$ based on information available by that date. In the applications
which we will deal with (capital growth), the standard models, e.g. \cite%
{AlgoetCover1988, Breiman1961, Kelly1956, Thorp1971}, exclude short selling.
Negative portfolio positions might lead to infinite negative values of
logarithmic functionals, playing a central role in the present context.
Following this approach, we will assume that all contingent portfolios $%
x_{t}(s^{t})$ are represented by non-negative vector functions. All
functions of $s^{t}$ will be assumed to be measurable and those representing
contingent portfolios essentially bounded.

Any sequence of contingent portfolios $%
x_{0}(s^{0}),x_{1}(s^{1}),x_{2}(s^{2}),...$ \ will be called a \textit{%
trading strategy}. Trading strategies describe possible scenarios of
investors' actions at the financial market influenced by random factors. In
the model, we are given sets $G_{t}(s^{t})\subseteq\mathbb{R}_{+}^{n}\times%
\mathbb{R}_{+}^{n}$ specifying the \textit{self-financing }(\textit{solvency}%
) constraints. The main focus of the study is on self-financing trading
strategies. A strategy $x_{0}(s^{0}),x_{1}(s^{1}),x_{2}(s^{2}),...$ is
called \textit{self-financing} if 
\begin{equation}
(x_{t-1}(s^{t-1}),x_{t}(s^{t}))\in G_{t}(s^{t})  \label{Z}
\end{equation}
almost surely (a.s.) for all $t\geq1$. The inclusion $%
(x_{t-1}(s^{t-1}),x_{t}(s^{t}))\in G_{t}(s^{t})$ means that the portfolio $%
x_{t-1}(s^{t-1})$ can be rebalanced to the portfolio $x_{t}(s^{t})$ at date $%
t$ in the random situation $s^{t}$ under transaction costs and trading
constraints. The rebalancing of a portfolio excludes inflow of external
funds, but it may take into account dividends paid by the assets.

It is assumed that for each $t\geq1$, the set $G_{t}(s^{t})$ is a \textit{%
closed convex cone} depending measurably\footnote{%
A closed set $G(s)\subseteq\mathbb{R}^{n}$ is said to depend measurably on a
parameter $s$ if the distance to this set from each point in $\mathbb{R}^{n}$
is a measurable function of $s$.} on $s^{t}$. This assumption means that the
model takes into account \textit{proportional} transaction costs. We give
examples of the cones $G_{t}(s^{t})$ below.

\textbf{Example 1.}\textit{\ No transaction costs.} Let 
\begin{equation*}
q_{t}(s^{t})=(q_{t}^{1}(s^{t}),...,q_{t}^{n}(s^{t})),\;q_{t}^{i}(s^{t})>0,
\end{equation*}
be the vector of the market prices of assets $i=1,2,...,n$ at date $t$.
Suppose that portfolio positions are measured in terms of the market values
of assets. Define 
\begin{equation}
G_{t}(s^{t}):=\{(a,b)\in\mathbb{R}_{+}^{n}\times\mathbb{R}_{+}^{n}:\sum
_{i=1}^{n}b^{i}\leq\sum_{i=1}^{n}\frac{q_{t}^{i}(s^{t})}{q_{t-1}^{i}(s^{t-1})%
}a^{i}\}.  \label{Z1}
\end{equation}
A portfolio $a=(a^{1},...,a^{n})$ can be rebalanced to a portfolio $%
b=(b^{1},...,b^{n})$ (without transaction costs) if and only if $(a,b)\in
G_{t}(s^{t})$.

\textbf{Example 2.}\textit{\ Proportional transaction costs: single currency.%
} Let $G_{t}(s^{t})$ be the set of those $(a,b)\in\mathbb{R}_{+}^{n}\times%
\mathbb{R}_{+}^{n}$ for which 
\begin{equation*}
\sum_{i=1}^{n}(1+\lambda_{t,i}^{+}(s^{t}))\,(b^{i}-\frac{q_{t}^{i}(s^{t})}{%
q_{t-1}^{i}(s^{t-1})}a^{i})_{+}\leq
\end{equation*}

\begin{equation}
\sum_{i=1}^{n}(1-\lambda_{t,i}^{-}(s^{t}))\,(\frac{q_{t}^{i}(s^{t})}{%
q_{t-1}^{i}(s^{t-1})}a^{i}-b^{i})_{+}\,,  \label{Z2}
\end{equation}
where $r_{+}:=\max\{r,0\}$ for the real number $r$. The transaction cost
rates for buying and selling are given by the numbers $%
\lambda_{t,i}^{+}(s^{t})\geq0$ and $1>\lambda_{t,i}^{-}(s^{t})\geq0$,
respectively. A portfolio $a=(a^{1},...,a^{n})$ can be rebalanced to a
portfolio $b=(b^{1},...,b^{n})$ (with transaction costs) if and only if the
pair of vectors $(a,b)$ belongs to the cone $G_{t}(s^{t})$. Here, we again
assume that the coordinates $a^{i}$ and $b^{i}$ of the portfolio vectors
indicate the current market values of the asset holdings. The inequality in (%
\ref{Z2}) expresses the fact that purchases of assets are made only at the
expense of sales of other assets. The approach based on relations (\ref{Z2})
is standard in the analysis of transaction costs; see e.g. Jouini and Kallal 
\cite{JouiniKallal1995}, Cvitani\'{c} and Karatzas \cite%
{CvitanicKaratzas1996}, and Pham and Touzi \cite{PhamTouzi1999}.

\textbf{Example 3.}\textit{\ Proportional transaction costs: several
currencies.} Consider an asset market where $n$ currencies are traded.
Suppose that for each $t=1,2,...$ a matrix

\begin{equation*}
\mu _{t}^{ij}(s^{t})\;\text{with }\mu _{t}^{ij}>0\;\text{and }\mu _{t}^{ii}=1
\end{equation*}%
is given, specifying the exchange rates of the currencies $i=1,2,...,n$
(including transaction costs). The number $\mu _{t}^{ij}(s^{t})$ shows how
many units of currency $i$ can be obtained by exchanging one unit of
currency $j$. A portfolio $a=(a^{1},...,a^{n})$ of currencies can be
exchanged to a portfolio $b=(b^{1},...,b^{n})$ at date $t$ in the random
situation $s^{t}$ if and only if there exists a nonnegative matrix $%
(d_{t}^{ji})$ (\textit{exchange matrix}) such that 
\begin{equation*}
a^{i}\geqslant \sum_{j=1}^{n}d_{t}^{ji},\;0\leqslant b^{i}\leqslant
\sum_{j=1}^{n}\mu _{t}^{ij}(s^{t})d_{t}^{ij}.
\end{equation*}%
Here, $d_{t}^{ij}$ ($i\neq j$) stands for the amount of currency $j$
exchanged into currency $i$. The amount $d_{t}^{ii}$ of currency $i$ is left
unexchanged. The second inequality says that at time $t$ the $i$th position
of the portfolio cannot be greater than the sum $\sum_{j=1}^{n}\mu
_{t}^{ij}d_{t}^{ij}$ obtained as a result of the exchange. The model we deal
with here is a version of the multicurrency models considered by Kabanov,
Stricker and others (see e.g. \cite{Kabanov1999}, \cite{KabanovStricker2001}
and \cite{Kabanov2001}). In spite of some similarity, it cannot be included
into the framework developed in the above papers. Note that in this example
asset holdings are expressed in terms \textquotedblleft physical
units\textquotedblright\ of assets (currencies).

An important class of dynamic securities market models is formed by \textit{%
stationary model}s. They are defined as follows. A model is called
stationary if the stochastic process $(s_{t})$ is stationary\footnote{%
Recall that a stochastic process $(s_{t})$ is called stationary if for any $%
m=1,2,...$ and any measurable function $\phi $ on the product of $m$ copies
of the space $S\times ...\times S$, the distribution of the random variable $%
\phi _{t}:=\phi (s_{t+1},...,s_{t+m})$ does not depend on $t$.} and the
given cones $G_{t}(s^{t})$ (specifying the solvency constraints) are of the
following form: 
\begin{equation}
G_{t}(s^{t})=G(s^{t}),  \label{ZG}
\end{equation}
where for each $s^{t}$ the set $G(s^{t})$ is a closed convex cone in $%
\mathbb{R}_{+}^{n}\times \mathbb{R}_{+}^{n}$ depending measurably on $s^{t}$%
. Assumption (\ref{ZG}) expresses the fact that the solvency constraints do
not explicitly depend on time: their structure depends only on the current
and previous states of the world---on the history $s^{t}$ of the underlying
stochastic process. In the stationary context it is convenient to assume
that $s_{t}$ is defined for each $t=0,\pm 1,\pm 2,...$, and in this case the
notation $s^{t}$ refers to the infinite history $s^{t}=(...,s_{t-1},s_{t})$.
This convention will always apply when we shall deal with stationary models.

If the stochastic process $(s_{t})$ is stationary, then the models
considered in Examples 1 and 2 are stationary if the asset returns $%
R_{t}(s^{t}):=q_{t}^{i}(s^{t})/q_{t-1}^{i}(s^{t-1})$ and the transaction
cost rates $\lambda_{t,i}^{-}(s^{t})$ and $\lambda_{t,i}^{+}(s^{t})$ do not
explicitly depend on $t$: 
\begin{equation*}
R_{t}(s^{t})=R(s^{t}),\;\lambda_{t,i}^{\pm}(s^{t})=\lambda_{i}^{\pm}(s^{t}).
\end{equation*}
The analogue of this assumption in the Example 3 is the condition that the
exchange rates do not explicitly depend on $t$: $\mu_{t}^{ij}(s^{t})=\mu
^{ij}(s^{t})$.

In the analysis of stationary models, we will consider a class of trading
strategies called \textit{balanced}. A strategy $x_{0},x_{1},x_{2},...$ is
termed balanced if there exist a vector function $x(s^{0})\in \mathbb{R}%
_{+}^{n}$ and scalar function $\alpha (s^{0})>0$ such that 
\begin{equation}
x_{0}(s^{0})=x(s^{0});\;x_{t}(s^{t})=\alpha (s^{t})...\alpha
(s^{1})x(s^{t}),\;t\geq 1,  \label{st2}
\end{equation}
and $|x(s^{0})|=1$. (We write $|\cdot |$ for the sum of the absolute values
of the coordinates of a vector). According to (\ref{st2}), portfolios $%
x_{t}(s^{t})$ grow with stationary proportions defined by the random vector
process $x(s^{0}),x(s^{1}),...$ and at a stationary rate $\alpha
(s^{1}),\alpha (s^{2}),...$. The results of capital growth theory pertaining
to stationary models (see Section 3) will be stated in terms of balanced
trading strategies.

\section{Von Neumann--Gale dynamical systems}

Von Neumann-Gale dynamical systems are defined in terms of multivalued
(set-valued) operators possessing properties of convexity and homogeneity.
States of such systems are represented by elements of convex cones $X_{t}$ ($%
t=0,1,...$) in linear spaces. Possible one-step transitions from one state
to another are described in terms of given operators $A_{t}(x)$, assigning
to each $x\in X_{t-1}$ a convex subset $A_{t}(x)\subseteq X_{t}$. It is
assumed that the graphs $Z_{t}:=\{(x,y)\in$ $X_{t-1}\times X_{t}:$ $y\in
A_{t}(x)\}$ of the operators $A_{t}(x)$ are convex cones. \textit{Paths} (%
\textit{trajectories}) of the von Neumann-Gale dynamical system are
sequences $x_{0}\in X_{0},x_{1}\in X_{1},...$ such that $x_{t}\in
A_{t}(x_{t-1})$.

In this work we consider stochastic von Neumann-Gale dynamical systems in
which a stochastic process $(s_{t})$ and a sequence of random closed convex
cones $G_{t}(s^{t})\subseteq \mathbb{R}_{+}^{n}\times \mathbb{R}_{+}^{n}$ ($%
t=1,2,...$) are given. The random elements $s_{t}$ of a measurable space $S$
are defined either for all non-negative integers $t$ or for all integers $t$%
. In the former case $s^{t}:=(s_{0},...,s_{t})$ and in the latter $%
s^{t}:=(...s_{t-1},s_{t})$. We denote by $\mathcal{X}_{t}$ the cone of
measurable essentially bounded vector functions $x(s^{t})$ with values in $%
\mathbb{R}_{+}^{n}$ and we put 
\begin{equation}
Z_{t}=\{(x,y)\in \mathcal{X}_{t-1}\times \mathcal{X}%
_{t}:(x(s^{t-1}),y(s^{t}))\in G_{t}(s^{t})\;\text{(a.s.)}\},\;  \label{Z5}
\end{equation}
\begin{equation}
A_{t}(x):=\{y\in \mathcal{X}_{t}:(x,y)\in Z_{t}\}.  \label{A}
\end{equation}
The multivalued operators $x\mapsto A_{t}(x)$ ($t=1,2,...$) transforming
elements of $\mathcal{X}_{t-1}\ $into subsets of $\mathcal{X}_{t}$ define
the von Neumann-Gale dynamical system we deal with. Paths of this system are
sequences of vector functions $x_{t}(s^{t})$ such that $x_{t}\in \mathcal{X}%
_{t}$ and $x_{t}\in A_{t}(x_{t-1})$. In the applications we have in mind,
these paths are self-financing investment strategies in the dynamic
securities market model described in the previous section and $G_{t}(s^{t})$
are the solvency cones in this model.

It is assumed that the cone $G_{t}(s^{t})$ depends measurably on $s^{t}$,
and for all $t$ the following basic conditions hold:

\noindent(\textbf{G.1}) for any $a\in$ $\mathbb{R}_{+}^{n}$, the set $%
\{b:(a,b)\in G_{t}\left( s^{t}\right) \}$ is non-empty;\medskip

\noindent(\textbf{G.2}) the set $G_{t}\left( s^{t}\right) $ is contained in $%
\{(a,b):|b|\leq M_{t}|a|\}$, where $M_{t}$ is a constant independent of $%
s^{t}$;\medskip

\noindent(\textbf{G.3}) there exist a strictly positive constant $\gamma
_{t}>0$ and a pair of essentially bounded vector functions $(\check{a}%
_{t-1}(s^{t}),\check{b}_{t}(s^{t}))$ such that $(\check{a}_{t-1}(s^{t}),%
\check{b}_{t}(s^{t}))\in G_{t}(s^{t})$ for all $s^{t}$ and $\check{b}%
_{t}(s^{t})\geq\gamma_{t}e$, where $e=(1,...,1)$;

\noindent(\textbf{G.4}) if $(a,b)\in G_{t}\left( s^{t}\right) $, $a^{\prime
}\geq a$ and $0\leq b^{\prime}\leq b$, then $(a,b)\in G_{t}\left(
s^{t}\right) $ (``free disposal hypothesis'').

All inequalities between vectors, strict and non-strict, are understood
coordinatewise.

Define 
\begin{equation}
G_{t}^{\times }(s^{t})=\{(c,d)\geq 0:db-ca\leq 0\;\text{for all }(a,b)\in
G_{t}(s^{t})\},  \label{G-dual}
\end{equation}%
where $ca$ and $db$ denote the scalar products of the vectors. Let $\mathcal{%
P}_{t}$ denote the set of measurable vector functions $p(s^{t})$ with values
in $\mathbb{R}_{+}^{n}$ such that $E|p(s^{t})|<\infty $. A \textit{dual path}
(\textit{dual trajectory}) is a finite or infinite sequence $%
p_{1}(s^{t}),p_{2}(s^{t}),...$ such that$\;p_{t}\in \mathcal{P}_{t}$ ($t\geq
1$) and 
\begin{equation}
(p_{t}(s^{t}),E_{t}p_{t+1}(s^{t}))\in G_{t}^{\times }(s^{t})\text{ (a.s.)}
\label{d2}
\end{equation}%
for all $t\geq 1$. We write $E_{t}(\cdot )=E(\cdot |s^{t})$ for the
conditional expectation given $s^{t}$. By virtue of (\ref{G-dual}) and (\ref%
{d2}), $E_{t}(p_{t+1}y)\leq p_{t}x$ (a.s.) for any $(x,y)\in Z_{t}$. This
inequality shows that for any path $x_{0},x_{1},...$ the sequence of random
variables $p_{1}x_{0},p_{2}x_{1},...$ is a supermartingale with respect to
the given filtration in the underlying probability space generated by $s^{t}$%
.

A dual path $p_{1},p_{2},...$ is said to \textit{support} a path $%
x_{0},x_{1},...$ if 
\begin{equation}
p_{t}x_{t-1}=1\;\text{(a.s.)}  \label{s1}
\end{equation}%
for all $t\geq 1$. A trajectory is called \textit{rapid} if there exists a
dual trajectory supporting it. The term \textquotedblleft
rapid\textquotedblright\ is motivated by the fact that 
\begin{equation*}
\frac{E_{t}(p_{t+1}y_{t})}{p_{t}y_{t-1}}\leq \frac{E_{t}(p_{t+1}x_{t})}{%
p_{t}x_{t-1}}=1\;\text{(a.s.)}
\end{equation*}%
for each path $y_{0},y_{1},...$ with $p_{t}y_{t-1}>0$ (see (\ref{d2}) and (%
\ref{s1})). This means that the path $x_{0},x_{1},...$ maximizes the
conditional expectation of the\textit{\ growth rate }at each time $t$, the
maximum being equal to $1$. Growth rates are measured in terms of the random
linear functions $p_{t}a$ of $a\in \mathbb{R}_{+}^{n}$. If states $x_{t}$ of
the von Neumann-Gale system represent portfolios whose positions are
expressed in terms of units of assets, then $p_{t}$ can be interpreted as
asset price vectors. If the $i$th coordinate $x_{t}^{i}$ of the vector $x_{t}
$ stands for the market value of the $i$th position of the portfolio, then $%
p_{t}^{i}$ may be regarded as a discount factor for the market price of the $%
i$th asset. Another motivation of the term \textquotedblleft rapid
path\textquotedblright\ lies in the fact that any rapid path is
asymptotically optimal---see the next section.

\section{Capital growth theory and von Neumann-Gale dynamical systems}

From the point of view of capital growth, those investment strategies are of
primary interest for which investor's wealth grows at an asymptotically
optimal rate. There are various approaches to the notion of asymptotic
optimality. In the definition below, we follow essentially Algoet and Cover 
\cite{AlgoetCover1988}.

\textbf{Definition 1. }Let $x_{0},x_{1},...$ be an investment strategy. It
is called \textit{asymptotically optimal} if for any other investment
strategy $y_{0},y_{1},...$ there exists a supermartingale $\xi _{t}$ such
that 
\begin{equation*}
\frac{|y_{t}|}{|x_{t}|}\leq \xi _{t},\;t=0,1,...\;\text{(a.s.).}
\end{equation*}

Recall that for a vector $b=(b^{1},...,b^{n})$ we write $%
|b|=|b^{1}|+...+|b^{n}|$. If $b\geq 0$, then $|b|=b^{1}+...+b^{n}$, and if
the vector $b$ represents a portfolio whose positions are measured in terms
of the market values of assets, then $|b|$ is the market value of this
portfolio. Note that the above property remains valid if $|b|$ is replaced
by any function $\psi _{t}(s^{t},b)$ (possibly random and depending on $t$)
which satisfies 
\begin{equation}
l|b|\leq \psi _{t}(s^{t},b)\leq L|a|,  \label{c-C}
\end{equation}
where $0<l<L$ are non-random constants. As an example of such a function, we
can consider the\textit{\ liquidation value }(or \textit{net asset value})
of the portfolio 
\begin{equation*}
\psi _{t}(s^{t},b)=\sum_{i=1}^{n}(1-\lambda _{t,i}^{-}(s^{t}))b^{i}
\end{equation*}
within the model defined by (\ref{Z2}). This is the amount of money the
investor gets if he/she decides to liquidate the portfolio (sell all the
assets) at date $t$. Clearly condition (\ref{c-C}) holds if the random
variables $1-\lambda _{t,i}^{-}>0$ are uniformly bounded away from zero.

The strength of the above definition, which might seem not immediately
intuitive, is illustrated by the following implications of asymptotic
optimality. As long as$\;|y_{t}|/|x_{t}|\leq\xi_{t}$,$\;t=0,1,...\;$(a.s.),
where $\xi_{t}$ is a supermartingale, the following properties hold.

(a) With probability one 
\begin{equation*}
\sup_{t}\frac{|y_{t}|}{|x_{t}|}<\infty\;\text{,}
\end{equation*}
i.e. for no strategy wealth can grow asymptotically faster than for $%
x_{0},x_{1},...$ (a.s.).

(b) The strategy $x_{0},x_{1},...$ a.s. maximizes the exponential growth
rate of wealth

\begin{equation*}
\lim\sup\limits_{t\rightarrow\infty}\frac{1}{t}\ln|x_{t}|.
\end{equation*}

(c) We have 
\begin{equation*}
\sup_{t}E\frac{|y_{t}|}{|x_{t}|}<\infty \;\text{\ and \ }\sup_{t}E\ln \frac{%
|y_{t}|}{|x_{t}|}<\infty .
\end{equation*}

Assertion (a) follows from a.s. convergence of non-negative
supermartingales; (b) is immediate from (a); the first part of (c) holds
because $\xi _{t}$ is a non-negative supermartingale; the second part of (c)
is obtained by using Jensen's inequality and the supermartingale property: $%
E(\ln \xi _{t+1}|s^{t})\leq \ln E(\xi _{t+1}|s^{t})\leq \ln \xi _{t}$.

This work aims at obtaining results on optimal growth in the model with
transaction costs described in Section 2. The main results are concerned
with the existence of asymptotically optimal strategies in the general
(non-stationary) version of the model and the existence of\textit{\ }%
asymptotically optimal balanced strategies in its stationary version. Our
main tool for analyzing the questions of asymptotic optimality is the
concept of a rapid path in the stochastic von Neumann-Gale system (see the
previous section).

\textbf{Definition 2.} A self-financing trading strategy $x_{0},x_{1},...$
is called \textit{rapid} if it forms a rapid path in the underlying von
Neumann-Gale dynamical system which defines the asset market model.

When dealing with the dynamic securities market model defined in terms of a
von Neumann-Gale dynamical system, we will use the terms \textquotedblleft
paths\textquotedblright\ and \textquotedblleft self-financing trading
strategies\textquotedblright\ interchangeably.

In the context of the present model, rapid paths may be regarded as
analogues of \textit{numeraire portfolios} (Long \cite{Long1990}). As we
have noticed, the price system (or the system of discount factors) $(p_{t})$
involved in the definition of a rapid path is such that the value $%
p_{t+1}x_{t}$ of the portfolio $x_{t}$ is always equal to one, while for any
other feasible sequence $(y_{t})$ of contingent portfolios (self-financing
trading strategy), the values $p_{t+1}y_{t}$ form a supermartingale. In the
classical case when transaction costs are absent, these conditions hold for
the price vectors $p_{t+1}:=\lambda _{t}q_{t}$, where $q_{t}$ are the market
prices and $\lambda _{t}^{-1}=q_{t}x_{t}$ is the market value of the
numeraire portfolio $x_{t}$. The latter is defined so that the normalized
prices $q_{t}/q_{t}x_{t}$ form a supermartingale. (Long \cite{Long1990}
considered a model with unlimited short selling, and in that context one can
speak of martingales rather than supermartingales.)

The results are based on assumption (\textbf{G.5}) below.

(\textbf{G.5}) There exists an integer $l\geq1$\ such that for every $t\geq0$
and $i=1,...,n$ there is a path $y_{t,i},...,y_{t+l,i}$\ over the time
interval $[t,t+l]$ satisfying 
\begin{equation*}
y_{t,i}=e_{i},...,y_{t+l,i}\geq\gamma e,
\end{equation*}
where $e_{i}=(0,0,...,1,...,0)$ (the $i$th coordinate is $1$ while the
others are $0$) and $\gamma$ is a strictly positive non-random constant.

\textbf{Proposition 1.} \textit{If the constants }$M_{t}$\textit{\ in
condition (\textbf{G.2}) do not depend on }$t$\textit{\ and assumption 
\textbf{(G.5) }holds, then any rapid path is asymptotically optimal.}

Thus in order to prove the existence of asymptotically optimal strategies it
is sufficient to establish the existence of infinite rapid paths. For a
proof of Proposition 1 see Evstigneev and Fl\aa m \cite{EvstigneevFlaam1998}%
, Proposition 2.5. In specific dynamic securities market models, condition (%
\textbf{G.5}) holds typically with $l=1$. Then it means a possibility of
buying some fixed strictly positive amounts of all the assets by selling one
unit of any asset $i=1,...,n$ (or if portfolio positions are measured in
terms of the market values of assets---by selling the amount of asset $i$
worth a unit of cash).

The main results of this paper are collected in the following theorem.

\textbf{Theorem 1.} \textit{(i) Let }$x_{0}(s^{0})$\textit{\ be a vector
function in }$\mathcal{X}_{0}$ \textit{such that }$ce\leq x_{0}(s^{0})\leq
Ce $\textit{\ for some constants }$0<c\leq C$\textit{. Then there exists an
infinite rapid path with initial state }$x_{0}(s^{0})$\textit{. (ii) If the
model is stationary and (\textbf{G.5}) holds, then there exists a balanced
rapid path. (iii) If the constants }$M_{t}$\textit{\ in condition (\textbf{%
G.2}) do not depend on }$t$\textit{\ and assumption \textbf{(G.5) }holds,
then the rapid paths whose existence is claimed in (i) and (ii) are
asymptotically optimal.}

Assertion (iii) is immediate from Proposition 1. Statement (i) of the above
theorem is proved in \cite{BahsounEvstigneevTaksar2008}, where the existence
of infinite rapid paths with the given initial state is established. The
proof in \cite{BahsounEvstigneevTaksar2008} is conducted by passing to the
limit from finite time horizons, for which the existence of rapid paths is
obtained in \cite{EvstigneevFlaam1998}. The passage to the limit is based on
a compactness principle involving Fatou's lemma in several dimensions
(Schmeidler \cite{Schmeidler1970}).

Assertion (ii) follows from the results of papers \cite%
{EvstigneevSchenkHoppe2007,EvstigneevSchenkHoppe2008}, where not only the
existence of a rapid path is proved, but also it is shown that there exists
a balanced rapid path supported by a dual trajectory with the following
special structure: 
\begin{equation}
p_{1}(s^{1})=p(s^{1}),\;p_{t}(s^{t})=\frac{p(s^{t})}{\alpha
(s^{t-1})...\alpha (s^{1})},\;t=2,3,...,  \label{d-b-path1}
\end{equation}
where $\alpha (s^{1})>0$ and $p(s^{1})\geq 0$ are scalar and vector
functions such that $E|p(s^{1})|<\infty $ (\textit{balanced dual trajectory}%
). The triplet of functions $\alpha (\cdot ),\,p(\cdot )$, $\,x(\cdot )$
involved in (\ref{st2}) and (\ref{d-b-path1}) is called a \textit{von
Neumann equilibrium}. It can be shown that if $\alpha (\cdot ),\,p(\cdot
),\,x(\cdot )$ is a von Neumann equilibrium, then the balanced trajectory
defined by (\ref{st2}) maximizes $E\ln \alpha $ among all such trajectories.
This means by definition that (\ref{st2}) is a \textit{von Neumann path}.
The existence of a von Neumann equilibrium established in \cite%
{EvstigneevSchenkHoppe2007,EvstigneevSchenkHoppe2008} is a deep result
solving a problem that remained open for more than three decades. In the
former of the two papers \cite%
{EvstigneevSchenkHoppe2007,EvstigneevSchenkHoppe2008}, a version of the
existence theorem for a von Neumann equilibrium is obtained which deals with
an extended model defined in terms of \textit{randomized paths}. In the
latter paper, the final result is derived by using the method of elimination
of randomization (Dvoretzky, Wald and Wolfowitz \cite%
{DvoretzkyWaldWolfowitz1950}).

To use Theorem 1 in specific models, one has to verify assumptions (\textbf{%
G.1})--(\textbf{G.5}) (note that (\textbf{G.3}) is a consequence of (\textbf{%
G.5}) with $l=1$). In Example 1, these conditions follow from the assumption
that the the asset returns $%
R_{t}^{i}(s^{t}):=q_{t}^{i}(s^{t})/q_{t-1}^{i}(s^{t-1})$ are uniformly
bounded and uniformly bounded away from zero. To obtain (\textbf{G.1})--(%
\textbf{G.5}) in Example 3 it is sufficient to assume that the exchange
rates $\mu _{t}^{ij}(s^{t})$ are uniformly bounded away from zero and
infinity. In Example 2, all the conditions needed can be obtained if the
above assumption regarding $R_{t}^{i}(s^{t})$ holds and the following
requirement regarding the transaction costs is fulfilled: the random
variables $\lambda _{t,i}^{+}(s^{t})$ are uniformly bounded and the random
variables $1-\lambda _{t,i}^{-}(s^{t})>0$ are uniformly bounded away from
zero. In all the three cases, (\textbf{G.2}) holds with constants $M_{t}$
independent of $t$.

\section{Conclusion}

This work deals with an important class of multivalued random dynamical
systems originally studied in mathematical economics---von Neumann--Gale
dynamical systems. We show how they can be applied to the analysis of some
fundamental issues in finance. This approach allows us to establish a link
with the classical von Neumann and Gale economic growth models, which makes
it possible to use concepts, techniques and results from mathematical
economics to obtain new theoretical results in finance. Even though one
would think that models of economic dynamics are the ``next of kin''\ to
dynamic security market models, surprisingly they have not been analyzed
from this angle for quite a while, and interconnections between these two
types of modeling frameworks have not been examined in as much detail as
they deserve. In a previous paper \cite{DempsterEvstigneevTaksar2006}, this
approach was applied to questions of asset pricing and hedging under
transaction costs and portfolio constraints. In the present study we show
how it can be employed to develop a theory of capital growth under
proportional transaction costs.\bigskip

\textbf{Acknowledgment.} Financial support from the Swiss National Center of
Competence in Research \textquotedblleft Financial Valuation and Risk
Management\textquotedblright\ (NCCR FINRISK), the grant NSF DMS-0505435, the
State of Missouri Research Board, the University of Missouri-Columbia
Research Council, and the Manchester School Visiting Fellowship Fund is
gratefully acknowledged.

\end{document}